# A review of the structure of street networks

Marc Barthelemy[1,2]

[1] Université Paris-Saclay, CNRS, CEA, Institut de Physique Théorique, Gif-sur-Yvette, France.
[2] Centre d'Analyse et de Mathématique Sociales CAMS, UMR 8557 CNRS-EHESS, Ecole des Hautes Etudes en Sciences Sociales, Paris, France

Geoff Boeing[3]

[3] Department of Urban Planning and Spatial Analysis, Sol Price School of Public Policy, University of Southern California, 301A Lewis Hall, Los Angeles, CA 90089-0626, USA

## Abstract

We review measures of street network structure proposed in the recent literature, establish their relevance to practice, and identify open challenges facing researchers. These measures' empirical values vary substantially across world regions and development eras, indicating street networks' geometric and topological heterogeneity.

## 1. Questions

Street networks shape the movement of people and goods through cities. Researchers model them to analyze network performance, predict traffic patterns, and evaluate infrastructure investments. Measures proposed in the recent network science literature alongside worldwide urban data availability can improve our understanding of these systems. How can we measure the spatial and topological structure of urban street networks to characterize the diversity of city organization worldwide?

## 2. Methods

Researchers usually model street networks as graphs where nodes are intersections and edges are street segments between intersections. This is the commonly accepted procedure, but it is not trivial, as discussed in (Marshall, et al., 2018). Data for contemporary streets and roads are often available from governmental sources as well as OpenStreetMap (OSM), a free, open worldwide geographic database updated and maintained by volunteers. Tools such as OSMnx (Boeing, 2017) render street network analysis fast and easy, using OSM data. We review papers from the past 20 years across the geography, engineering, planning, and physics literatures that propose structural measures. Then we compare their empirical findings around the world.

## 3. Results

Various properties of these networks were studied over the past two decades, e.g. (Cardillo, et al., 2006; Buhl, et al., 2006; Xie & Levinson, 2007; Crucitti, et al., 2006; Barthelemy & Flammini, 2008; Lämmer, et al., 2006; Barthelemy, et al., 2013; Strano, et al., 2012; Louf & Barthelemy, 2014; Levinson, 2012 ; Kirkley, et al., 2018; Chen, et al., 2024; Gudmundsson & Mohajeri, 2013; Jiang, 2007; Scellato, et al., 2006; Masucci, et al., 2009; Rosvall, et al., 2005), and here we will focus on the most important indicators. The temporal evolution of these structures is also studied, thanks to the digitalization of historical data (Perret, et al., 2015; Strano, et al., 2012; Masucci, et al., 2013; Barthelemy, et al., 2013; Burghardt, et al., 2022; Barrington-Leigh & Millard-Ball, 2015).

Due to strong physical constraints, many quantities studied in complex networks (Latora, et al., 2017; Menczer, et al., 2020) are not relevant for street networks (Lämmer, et al., 2006; Mossa, et al., 2002). For example, street networks have a narrow degree distribution and a very high clustering coefficient (Barthelemy, 2022). Also, many indicators introduced in transportation geography, such as alpha, beta, and gamma indices (Kansky, 1963), mainly depend on the average degree and are redundant. To effectively characterize street networks, it is also crucial to consider spatial properties such as planarity, road segment length distribution, betweenness centrality spatial distribution, block shape factor, and street angle distribution. We will present here results for what could constitute a minimal set of measures that characterizes a street network.

The most basic indicator, the number of nodes, increases with the population - which makes theoretical sense as residents can share public infrastructure - at most linearly (Strano, et al., 2012; Barthelemy, et al., 2013), and in general slightly sublinearly (Boeing, 2022): a 1% increase in urban area population is associated with a 0.95% (±0.03%) increase in intersection count. The total length scales accordingly to a simple argument (Barthelemy & Flammini, 2008) as $\sqrt{AN}$ where $A$ is the area size. Basic measures (see Supp. Info. for the definition of all quantities discussed) which convey important information about these networks are summarized in table 1 for 100 world cities.

| Quantity | $\overline{k}$ | $p_1$ | $p_4$ | $l_1$ |
|---|---|---|---|---|
| range | [2.30,3.55] | 2%-39% | 4%-59% | [64.5,537.5] |
| median | 2.93 | 14% | 18% | 118.6 |

Table 1. Basic network statistics computed for the set of 100 world cities discussed in (Boeing, 2019): average degree $\overline{k}$, proportion of dead-ends $p_1$, proportion $p_4$ of $k = 4$ intersections, average length $l_1$ (in meters) of edges.

We observe for all measures a large diversity of values and illustrate the possible networks according to their average degree value in Figure 1.

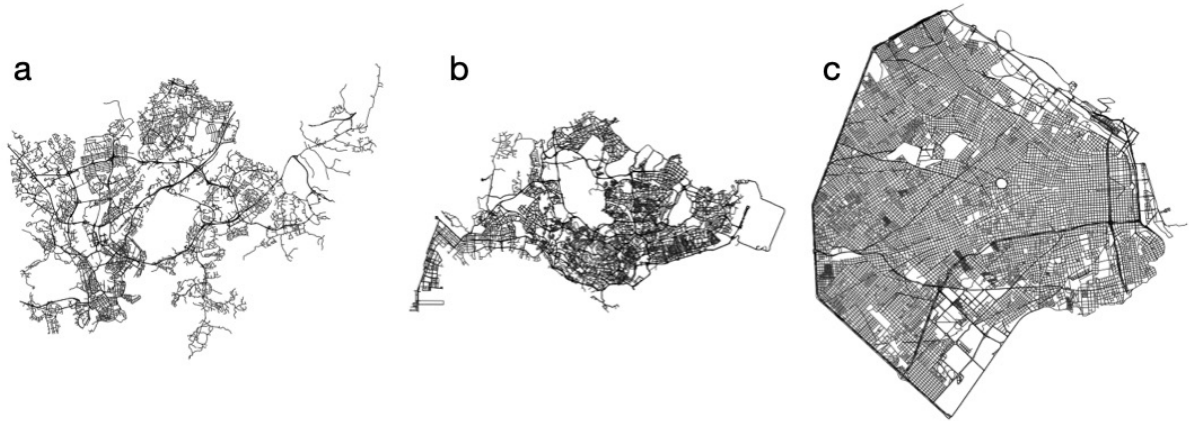

Figure 1. Street networks with (**a**) small average degree (Helsinki $\overline{k} = 2.35$), (**b**) typical value (Singapore $\overline{k} = 3.0$), and (**c**) a large value (Buenos Aires $\overline{k} = 3.55$). The overall organization of street networks is diverse, reflecting a wide range of patterns and structures depending on urban planning, geography, and cultural factors (see the Supp. Info. for details about the degree $k$ calculation).

These networks are not completely planar due to the presence of tunnels and bridges. A measure, the Spatial Planarity Ratio, $\varphi$ was introduced in (Boeing, 2020): a spatially planar network with no overpasses or underpasses will have $\varphi = 1.0$, while lower values indicate the extent to which the network is planar. Among drivable street networks for 50 world cities, only 20% are formally planar and on average $\varphi = 0.88$, and ranging from 100% in six of these cities to a low of 54% for Moscow.

T he betweenness centrality (BC) defined in (Freeman, 1977) measures the importance of a node (or edge) for flows on the network. In this sense it could serve as a simple proxy for traffic on the network (although it assumes in general a flat OD matrix), but also as an interesting structural probe of the network. The distribution of betweenness centrality (BC) is invariant for street networks, despite the existence of structural differences between them (Kirkley, et al., 2018). For a regular network, the BC decreases with the distance to the gravity center of nodes, but when disorder is present, we observe the emergence of different patterns. In particular, we observe the presence of loops with large BC (Lion & Barthelemy, 2017; Barthelemy, et al., 2013) signaling the importance of these structures for large cities (see figure 2).

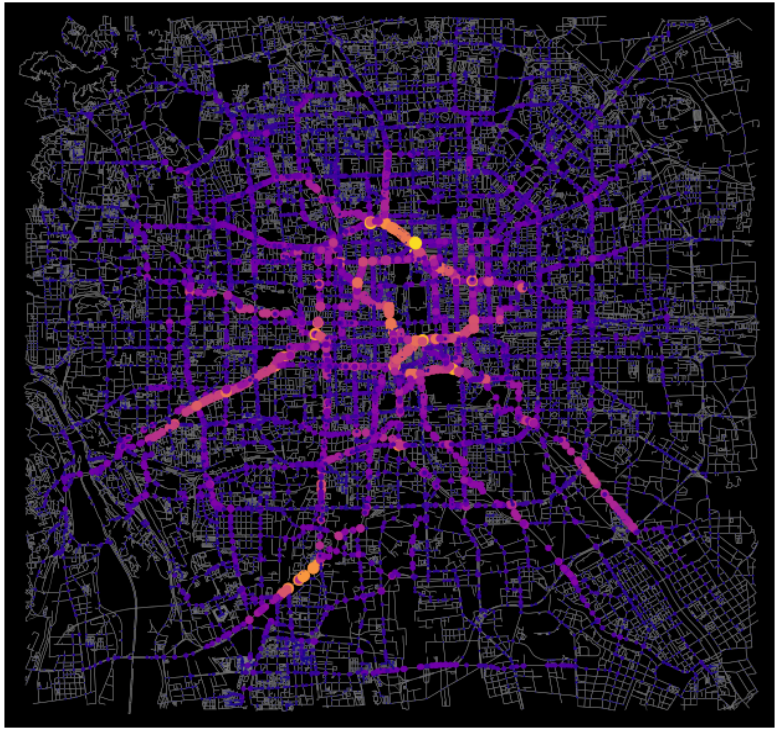

Figure 2. Spatial distribution of node betweenness centrality (weighted by edge length) for Beijing, China (yellow nodes are more central). We observe the emergence of non-trivial patterns of large BC nodes (Data from OSM).

The geometry of street networks is also fundamental. Using a large global database comprising all major roads on the Earth, (Strano, et al., 2017) showed that the road length distribution within croplands is indistinguishable from urban ones, once rescaled by the average road length. The area $a$ of blocks is another important feature of these spatial networks, and it was shown that its distribution is universal of the form $P(a) \sim a^{-2}$ (Lämmer, et al., 2006; Louf & Barthelemy, 2014). The organization and overall geometry of the street network can also be characterized by the distribution of street angles. There are very regular networks (almost lattice like) such as in Chicago and very disordered ones such as in Rome (Figure 3). The orientation order can then be characterized by the entropy of street compass bearings (Boeing, 2019).

Finally, several authors proposed to construct a typology of these street networks: Marshall (Marshall, 2004) proposed a first approach based on general considerations, a typology based on the block size distribution and shape was proposed in (Louf & Barthelemy, 2014), and more recently machine learning approaches were proposed. In (Thompson, et al., 2020) a greater proportion of railed public transport networks combined with dense road networks characterised by smaller blocks is correlated with the lowest rates of road traffic injury, and in (Boeing, et al., 2024) it was shown that straighter, more-connected, and less-overbuilt street networks are associated with lower transport emissions, all else equal.

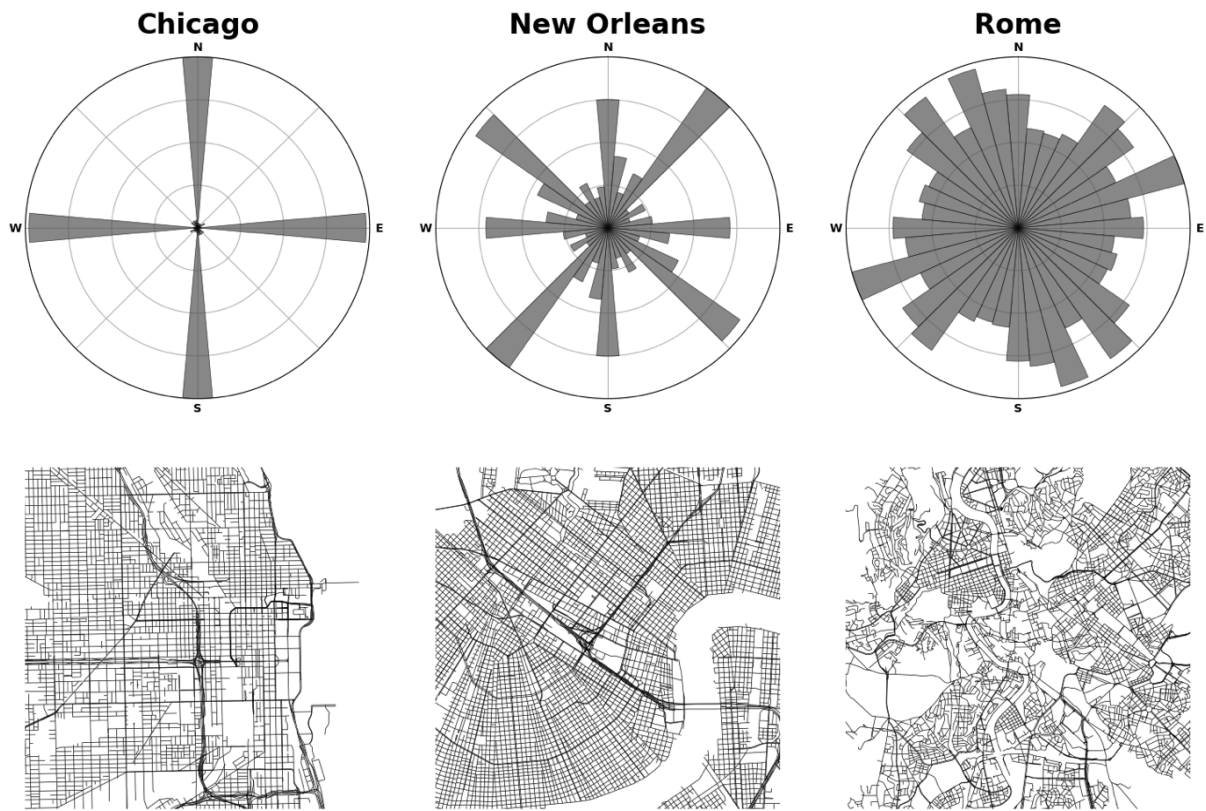

Figure 3. Polar histograms (bar directions represent streets' compass bearings and bar lengths represent relative frequency of streets) and the corresponding street maps illustrating different types of organization: low entropy (Chicago), typical (New Orleans), and high entropy (Rome).

This review covers a minimal set of topological and spatial measures of street network structure: node count, average degree, fraction of dead ends and intersections, average edge length, planarity index, BC spatial distribution, street angle distribution. Despite the universality of street networks, these measures show the diversity of structural patterns of streets reflecting local culture, politics, era, and transport technologies. Important challenges remain open. First, we need a better understanding of the spatio-temporal evolution of these networks and their co-evolution with cities (Capel-Timms, et al., 2024). More efforts in the digitization of historical maps are needed to advance our theoretical understanding and modeling of this phenomenon. Second, merely identifying empirical values is not enough for the professional disciplines of city-making. A key future challenge is to link the "what" (descriptive measures) to the "how" to build new and improve existing networks to meet broader societal goals (sustainability, resilience, public health, economic health, etc).

# Supplementary Information
# A review of the structure of street networks

Marc Barthelemy[1,2]

[1] Université Paris-Saclay, CNRS, CEA, Institut de Physique Théorique, Gif-sur-Yvette, France.
[2] Centre d'Analyse et de Mathématique Sociales CAMS, UMR 8557 CNRS-EHESS, Ecole des Hautes Etudes en Sciences Sociales, Paris, France

Geoff Boeing[3]

[3] Department of Urban Planning and Spatial Analysis, Sol Price School of Public Policy, University of Southern California, 301A Lewis Hall, Los Angeles, CA 90089-0626, USA

## Definitions

### 1. Degree

The street network is described by a network $G = (V, E)$ where $V$ is a set of $N$ nodes and $E$ the set of links between these nodes. The nodes represent the intersections and the links segments of roads between these intersections. The degree $k$ of a node is the number of streets converging to it. A node of degree $k = 1$ is a dead-end, nodes of degree $k = 2$ are generally removed and nodes of degree 3, 4. (or more) represent typical intersections. The average degree is then simply given by

$$\bar{k} = \frac{1}{N}\sum_{i=1}^{N} k_i$$

where $k_i$ is the degree of node $i$. In general, the number of nodes of degree $k$ is denoted by $N(k)$ and the proportion of dead-ends reads then

$$p_1 = \frac{N(1)}{N}$$

and of $k = 4$ intersections:

$$p_4 = \frac{N(4)}{N}$$

### 2. Detour index

The detour index (or stretch factor) for a pair of nodes $i$ and $j$ is defined as (Aldous & Shun, 2010; Barthelemy, 2022)

$$Q(i,j) = \frac{d_r(i,j)}{d_e(i,j)}$$

where $d_e$ is the Euclidean distance between $i$ and $j$, and $d_r$ is the route distance computed on the network. We then have

$$Q_{max} = \max_{i,j} Q(i,j)$$

We can also average over pairs of nodes at a given distance $d$ and construct the detour profile

$$Q(d) = \frac{1}{E(d)} \sum_{i,j \text{ s.t. } d_e(i,j)=d} Q(i,j)$$

where $E(d)$ is the number of pairs of nodes at distance $d$.

### 3. Total and average length

The total length of the network is defined as (Barthelemy, 2022)

$$L = \sum_{e \in E} l(e)$$

total length $l(e)$ is the length of edge $e$. The average edge length is then

$$l_1 = \frac{1}{N} L$$

### 4. Spatial planarity ratio

The spatial Planarity Ratio, $\varphi$ (Boeing, 2020) represents the ratio of the number $i_n$ of nonplanar intersections (i.e., non-dead-end nodes in the nonplanar, three-dimensional, spatially-embedded graph) to the number $i_p$ of planar intersections (i.e., edge crossings in the planar, two-dimensional, spatially-embedded graph):

$$\varphi = \frac{i_p}{i_n}$$

The (positive) quantity $i_p - i_n$ is then equal to the number of nonplanar edge crossings such as overpasses and underpasses in the network.

### 5. Fraction of one-way streets

The fraction $p$ of one-way streets is defined as (Verbavatz & Barthelemy, 2021)

$$p = \frac{L_1}{L}$$

where $L_1$ is the total length of one-way streets and $L$ the total length of the network.

### 6. Betweenness centrality

An interesting quantity, first discussed in the context of non-spatial network (Freeman, 1977) is the betweenness centrality (BC). The betweenness centrality (BC) of a node $i$ is defined as (Freeman, 1977)

$$g(i) = \frac{1}{(N-1)(N-2)} \sum_{s,t \neq i} \frac{\sigma_i(s,t)}{\sigma(s,t)}$$

where $\sigma(s,t)$ is the number of shortest paths from $s$ to $t$, and $\sigma_i(s,t)$ is the number of such shortest path that go through the node $i$ (and a similar definition for the BC of edges). The normalization (here chosen as the number of pairs of nodes different from $i$) can be slightly different according to different authors.